\documentclass[%
preprint,
nofootinbib,
amsmath,amssymb,
]{revtex4}

\usepackage{graphicx}% Include figure files
\usepackage{dcolumn}% Align table columns on decimal point
\usepackage{bm}% bold math
\begin{document}

\preprint{CYCU-HEP-19-07}

\title{Hawking Radiation As Stimulated Emission}%

\author{Wen-Yu Wen}\thanks{wenw@cycu.edu.tw}
 \affiliation{Department of Physics and Center for Theoretical Physics,\\
 Chung Yuan Christian University, Taiwan}

\begin{abstract}
We regard the Parikh-Wilczek's tunnelling model of Hawking radiation as a quantum mechanical process of stimulated emission.  The hypothesized microstates are found at the horizon with double degeneracy.  A Jaynes-Cummings toy model for a black hole in the cavity is proposed to demonstrate how to write a {\sl qubit} via the angular-dependent transition coupling, which might be related to the soft Goldstone hairs at analytic continuation.  At last, we show how information is retained in the black hole by computing the time evolution of mutual entanglement entropy in the cavity-black holes system.  
\end{abstract}

\keywords{Hawking Radiation, Jaynes-Cummings Model, Stimulated Emission}

\maketitle

%\tableofcontents

\section{Introduction and Summary}

\begin{figure}[t]
\includegraphics{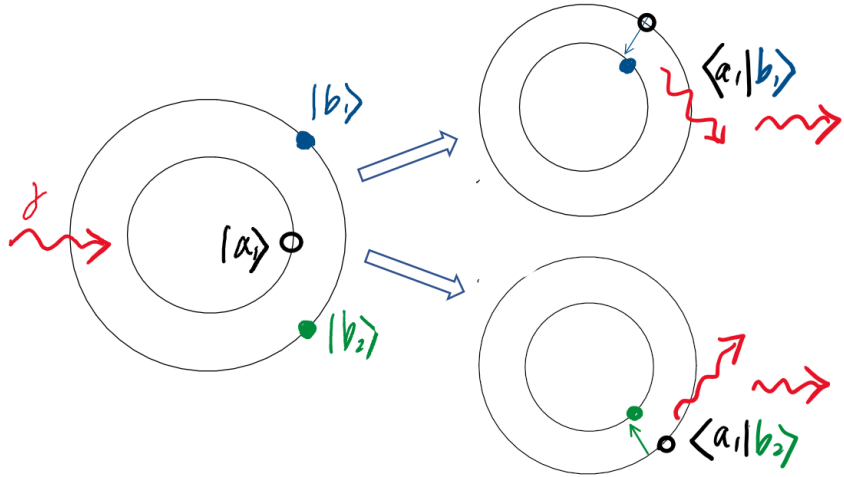}
\caption{\label{fig:stimulated} (Left) Degenerate excited states $|b_1\big \rangle$ and $|b_2\big \rangle$ are stimulated by a photon in the cavity.  (Right) The stimulated emission may have different feature depending on which transition $\big\langle a_1 | b_1\big\rangle$ or $\big\langle a_1 | b_2\big\rangle$ occurs.}
\end{figure}

Parikh and Wilczek earlier gave a semiclassical derivation of Hawking radiation as a tunneling process, similar to pair creation in a constant electric field \cite{Parikh:1999mf}.  If one considers a particle with energy $\omega$ is emitted from a black hole with mass $M$. The emission rate reads\footnote{Here the natural unit is adopted that $\hbar = c = 1$　and Boltzmann constant $k_B=1$.  If causing no confusion, we also set $G=1$, or equivalently the Planck length $\l_p =1$ for convenience.}
\begin{equation}
\Gamma \sim e^{-8\pi\omega(M-\omega/2)} = \underbrace{e^{-\omega/T_H}}_{\textrm{Boltzmann factor}} {e^{+4\pi\omega^2}}.  
\end{equation}
It is impressive a nonthermal spectrum can be obtained by assuming conservation of energy during the tunneling process.  In addition, the exponent happens to be the very change of entropy, conservation of information is therefore achieved in terms of mutual information \cite{Zhang:2009jn}.  Nevertheless, it is unclear whether the infamous paradox of lost information can be resolved in this macroscopic picture since it still cannot reveal those hidden microstates responsible for the black hole entropy.  In this letter, we model the black hole as an {\sl atom} with many degenerate states and regard the Hawking radiation as a quantum mechanical process of stimulated emission \cite{Einstein:1916}.  Our quantum mechanical model of black holes has following features:
\begin{itemize}
    \item We argue that transition from different degenerate excited states maybe responsible for the featured radiation as depicted in the Fig. \ref{fig:stimulated}.  In the section II, we explicitly show that at large black hole limit, the nonthermal spectrum in the Parikh-Wilczek tunneling model can be reproduce.  This agreement implies the hypothesized microstates with double degeneracy were seated at the horizon.  It is also tempted to guess they are closed related to the soft gravitons which claim the black hole information.
    
    \item To justify our statement, in the section III we propose a Jaynes-Cummings (JC) toy model for a black hole in the cavity and demonstrate how to write a {\sl qubit} via the angular-dependent transition coupling.  In the section IV, we further argue that after analytic continuation, the coupling strength is related to the soft hairs, which is represented by the Goldstone boson modes. 
    
    \item The information stored in a black hole is believed to be carried by entangled microstates.  On the other hand, coherent states are expected to be dephased sooner or later in an open system via interaction.  It is curious how those information carriers survive long enough before they are given away to the featured emission.  In the section V, we show how entanglement among the excited microstates can be sustained for a while by computing the time evolution of mutual entanglement entropy in the cavity-black holes system.  Our result suggests the quantum information stored in a realistic black hole can last long as if it is {\sl preserved} in an invisible cavity bounded by the photon sphere.
\end{itemize}

\section{{\sl Black} Atoms In Cavity}

We only focus on the Schwarzschild black hole in this letter.  To avoid its runaway thermal misbehavior thanks to the negative specific heat, we consider a gedanken experiment as follows: one prepares a collection of many large Schwarzschild black holes confined in a much bigger but finite cavity with reflective walls.  We wait for some time until the photons emitted by the Hawking radiation are balanced by the absorbed photons bounced off the walls.  The cavity-black holes system eventually reaches thermal equilibrium at the Hawking temperature $T_H$.  This setup is in contrast to that in the tunneling scenario, for energy conservation is not enforced here.  We now regard our system as a collection of {\sl ${\cal N}$-level atoms} in the Einstein's quantum mechanical model of emission, where ${\cal N}$ could be as large as the black hole mass in the unit of Planck mass, say ${\cal O}(M/m_{pl})$.  For simplicity, we only consider the transition between two nearby levels.  The lower(upper) state $| a(b) \big \rangle$ corresponds to the quantum state of black hole with mass $M_{a(b)}$ after(before) radiation.  According to the common notation, let $A$ be the spontaneous emission rate and $B_{ba}\rho(\omega)$ be the transition rate for stimulated emission.  We have following additional assumptions:
\begin{itemize}
    \item Each state has $g_{a(b)}$ degeneracy, which should be exponentially large enough to accommodate degrees of freedom in a black hole.
    \item Those degrees of freedom are located somewhere at or outside the event horizon.
\end{itemize}
  Applying the detailed balancing to the cavity-black holes system, the radiation spectrum reads
\begin{equation}\label{eq:radiation_1}
   \rho(\omega) = \frac{A}{(N_a/N_b)B_{ab}-B_{ba}}
\end{equation}
In thermal equilibrium at Hawking temperature $T_{H}$, the black hole population in the state $\psi_{a(b)}$ is given by $N_{a(b)}=g_{a(b)}e^{-M_{a(b)}/T_H}$.  To have simple Maxwell-Boltzmann distribution, one would have demanded the relation $g_a B_{ab} = g_b B_{ba}$.  In our model, however, we have assumed $g_{a(b)} \sim e^{\frac{\alpha}{4} {\cal A}(\beta M_{a(b)})}$ in order to respect the area law such that it can host black hole's Bekenstein-Hawking entropy.  To be precise, we assume the hypothesized microstates are located at radius $r=\beta M_{a(b)}$ and ${\cal A}(\beta M_{a(b)})$ is the area of sphere with that radius in the isotropic coordinate.  Here the isotropic metric is adopted to honestly reflect spatial distance in sphere:
\begin{equation}
ds^2=-\frac{(1-M/2r)^2}{(1+M/2r)^2}dt^2+(1+M/2r)^4(dr^2+r^2d\Omega^2).
\end{equation}
The proportionality coefficients $\alpha$ and $\beta$ will be determined shortly. We are looking for large black hole limit where $1/M \ll \omega \ll M$, then equation (\ref{eq:radiation_1}) can be cast into
\begin{equation}
\rho(\omega) \simeq (A/B_{ba})e^{-\omega /T_H} e^{\alpha\frac{\pi}{4} C(M,\omega)},
\end{equation}
\begin{widetext}
where
\begin{equation}
C(M,\omega) =  (\frac{2}{\beta^2}+\frac{8}{\beta}-32\beta-32\beta^2)M\omega + (\frac{3}{\beta^2}+\frac{8}{\beta}+16\beta^2)\omega^2 + {\cal O}(\omega^3).
\end{equation}
\end{widetext}
We remark the choices for coefficients $\alpha$ and $\beta$ as follows:
\begin{itemize}
    \item To recover the Boltzmann factor, we choose $\beta = 1/2$ such that the leading term in function $C(M,\omega)$ vanishes.  This suggests those degrees of freedom are seated at the horizon \footnote{We note that in the isotropic coordinate, the horizon locates at $r=M/2$.}.
    \item To reproduce the Parikh-Wilczek nonthermal spectrum, we further choose $\alpha = 2$.  This implies that the degeneracy at each energy level is twice amount of the black hole entropy, for $S_{BH}=A/4$. 
\end{itemize}
It has been conjectured that black hole microstates are given by those soft hairs which enjoy the Bondi-Metzner-Sachs (BMS) transformation of supertrasnlation and superrotation \cite{Bondi:1962,Sachs:1962zza,Sachs:1962,Dvali:2011aa,Dvali:2015ywa,Dvali:2015rea, Hawking:2016msc, Strominger:2017aeh, Chu:2018tzu}.  This infinite dimensional global symmetry is expected to be broken by the horizon geometry to form a finite number of gapless Bogoliubov-Goldstone modes \cite{Averin:2016hhm, Eling:2016xlx}.  Our Einstein model of stimulated emission happens to agree with that picture that double at the horizon could be interpreted as $N$ massless spin-$2$ soft gravitons at criticality, where $N\sim e^{M^2/\l_p^2}$ respecting the area law.  Those microstates in coordinate representation can be modeled as {\sl Boolean} qubit at the lattice of Planckian spacing \cite{Sorkin:2014kta, tHooft:1993dmi} or in momentum representation as eigenstates labeled by angular quantum numbers, say $|\l,m,s \big \rangle$, where $s=\pm 2$, $|m|<\l$ and $l \le \l _{max} \sim \sqrt{N}$ \cite{Averin:2016ybl}. 

\section{Jaynes-Cummings Model of Black Atom}

We now construct a Jaynes-Cummings (JC) toy model for the black hole atom fit to our thought experiment.  The JC model was to study the interaction of a two-level atom with a single electromagnetic field \cite{Jaynes:1963}.  Now we would like to replace the atom by a black hole.  It could be dangerous to model a black hole without its spacetime context.  However, it has been argued that Hawking radiation, including the nonthermal feature, could be well formulated without spacetime \cite{Braunstein:2011gz}.  Therefore, our quantum mechanical model may be just enough to serve our purpose to explain this featured radiation.  After rotating-wave approximation, the JC Hamiltonian, which is composed of energy of black holes, radiation energy and interaction, reads
 \begin{equation}\label{eqn:hamiltonian}
  {\cal H} = \underbrace{ \sum_i M_a |a_i\big \rangle \big \langle a_i| + \sum_j M_b |b_j\big \rangle \big \langle b_j| }_{\textrm{black holes energy}}
  + \underbrace{ \omega \hat{\alpha}^{\dagger}\hat{\alpha} }_{\textrm{photon energy}}
  + \underbrace{ \sum_i \sum_j  g_{ij} \big( \hat{\alpha}|b_j\big \rangle \big \langle a_i| + \hat{\alpha}^{\dagger} |a_i\big \rangle \big \langle b_j| \big) }_{\textrm{interaction}}
 \end{equation}
where indices $i,j$ label each degeneracy state.  $\hat{\alpha}$ and $\hat{\alpha}^{\dagger}$ are annihilation and creation operators of photons.  The couplings $g_{ij}$ are responsible for emission and absorption.  Without loss of generality, we only focus on transition among just few states, say $|n,a_1\big \rangle$ for the ground state, and $|n-1,b_1\big \rangle$, $|n-1,b_2\big \rangle$ for two degenerate excited states.  We define the product state, for instance, $|n,a_1\big \rangle = |n\big \rangle \otimes |a_1\big \rangle$, where $n$ is the photon occupation number in the cavity and spin index $s$ is suppressed in atom states.  The eigenvalues and eigenstates are calculated in the Appledix.  In particular, at the resonance $\Delta=0$ and even coupling strength $g_{11}=g_{12}=g$, we have following remarks:
\begin{itemize}
    \item After analytic continuation of coupling $g\to ig$, the transition (\ref{eqn:time_evolution}) can be regarded as the Bogoliubov transformation between in state $\big|\cdot\big\rangle_0$ and out state $\big|\cdot\big\rangle_t$ similar to that in the parametric amplifier \cite{Clerk:2010} or a harmonic oscillator with upside-down potential.  The {\sl Rabi} oscillation frequency $\Omega\sim {\cal O}(g)$ after analytic continuation can be related to the Unruh temperature and therefore the coupling $g \sim 1/M$, which also agrees with the gravitational self-coupling strength at criticality in \cite{Averin:2016ybl}.
    
    \item The hopping probability among those states of same energy level, say $|n-1,b_1\big \rangle$ and $|n-1,b_2\big \rangle$, may represent the scramble rate, that is $\big|_t \big\langle n-1,b_2 | n-1,b_1 \big\rangle_0 \big|^2 = \big|-\frac{1}{2\sqrt{2}}+\frac{1}{2}\cos \Omega t \big|^2$.  This scramble is somehow triggered by interaction with resonant photons through coupling $g$.  In general, one may also introduce additional {\sl hopping} coupling among degenerate states, say $J\Big(|n-1,b_1\big \rangle \big \langle n-1,b_2 |+|n-1,b_2\big \rangle \big \langle n-1,b_1 |\Big)$ in the Hamiltonian (\ref{eqn:hamiltonian}).  This is about to shift the eigenvalue by some constant proportional to $J$ as well as $\Omega \to \sqrt{2g^2n+J^2}$.
    
    \item To account for the large degeneracy in a black hole,  the amount of various couplings $g_{ij}$ is of order $\sim {\cal O}(e^{M^2})$.  Therefore the Rabi frequency will be enhanced by the same power, which leads to fast scramble.
    
    \item In the case of uneven coupling strength $g_{11}\neq g_{12}$, the transition amplitude to $\big |n-1,b_1\big\rangle$ and $\big |n-1,b_2\big\rangle$ are  different.  This could be responsible for featured emission and therefore encoding information.  In terms of the BMS transformation of soft hairs, the variation in photon coupling $g_{ij}$ and hopping coupling $J_{ij}$ can be realized in some arbitrary angle-dependent function $F(\nu,\theta,\phi)$ \cite{Averin:2016hhm, Maitra:2019eix}.
    
    \item It was argued in \cite{Averin:2016ybl} that the degeneracy of above-mentioned Goldstone modes will be lifted by quantum fluctuation.  That is, one expects nontrivial tidal force due to gravitational perturbation, like the gravity version of Zeeman effect, contributes to a small energy gap \footnote{This small energy gap should not be confused with those among $E_0$ and $E_{\pm}$, which are induced by interaction with photons.}.   We may estimate its order by considering fluctuation in Newton's gravitational potential $ \delta U \sim \omega \delta g_{00} \sim M\omega\delta r/r^2 |_{r=2M}$.  Given the quantized radiation $\omega \sim \hbar/4M$ \cite{Wen:2014xpa} and $\delta r \sim \omega$, one obtains the energy gap $\Delta E \sim \delta U \sim \hbar /M^3 $ in agreement with \cite{Averin:2016ybl}.  However, its contribution to $\Omega$ is minor due to its small magnitude.

\end{itemize}

\begin{figure}[t]
\includegraphics[scale=0.45]{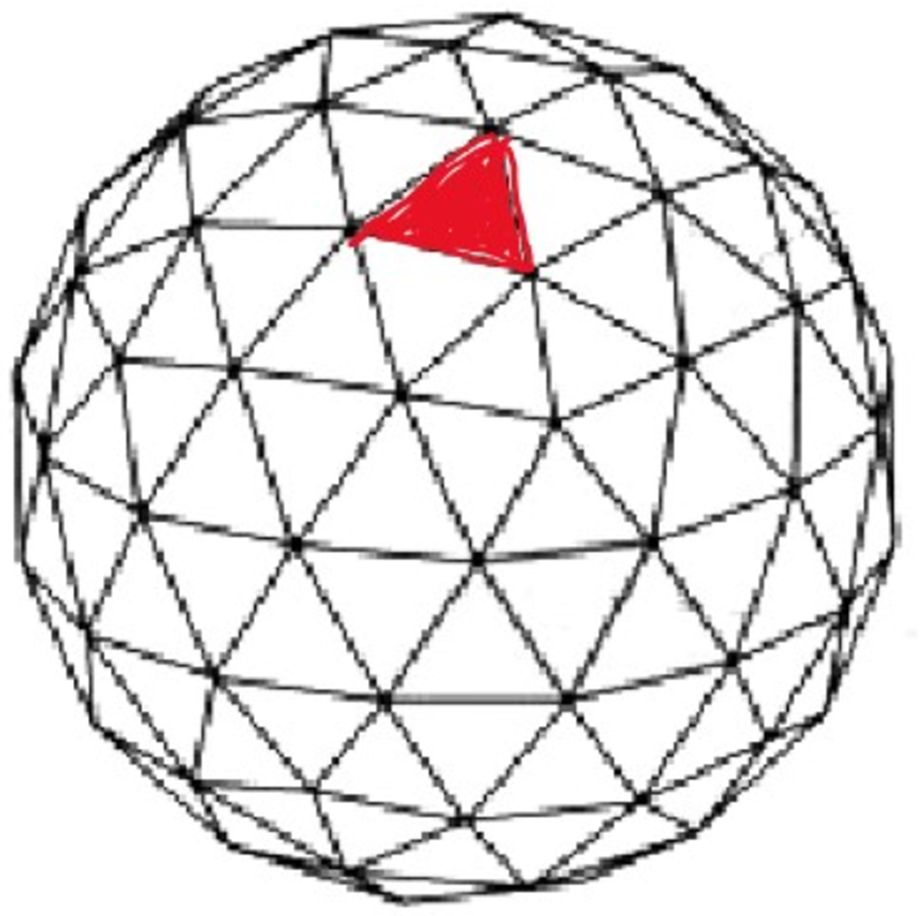}
\includegraphics[scale=0.45]{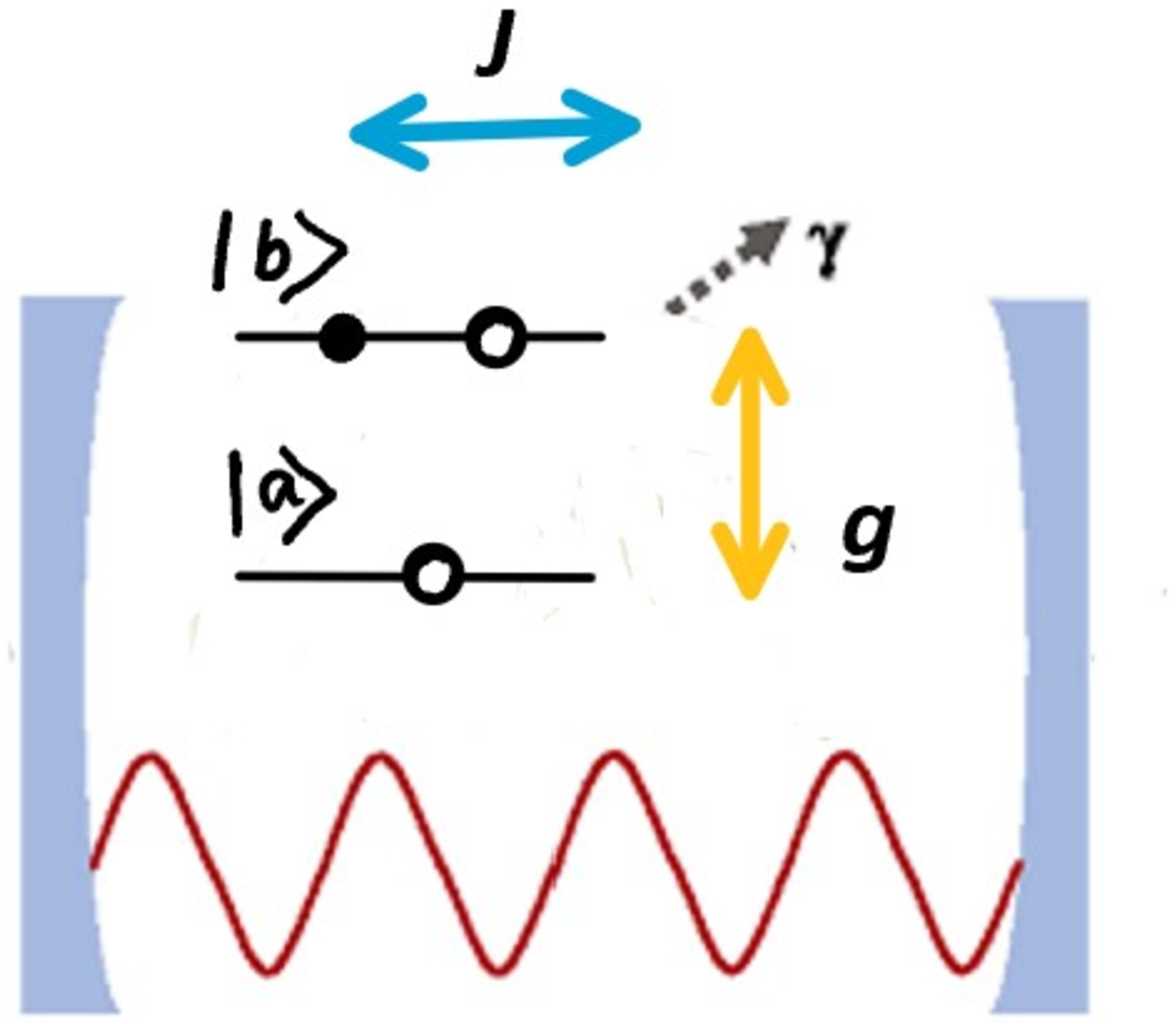}
\caption{\label{fig:JC model}(Left) Degenerate microstates of Planck size are seated at horizon. (Right) A black hole in the cavity is modeled by a two-level atom in the Jaynes-Cummings model.  Coupling $g$ is the transition strength between excited state $|b\big\rangle$ and ground state $|a\big\rangle$, while $J$ is the hopping strength among degenerate excited states.}
\end{figure}

\section{Soft-Hair Dressed Coupling Strength}
We have argued that uneven coupling strength could be responsible for featured emission, from which one might be able to decode the information.  While processing information in quantum mechanics is reversible thanks to unitarity property, we now show how to create a desired superposition state in the JC model by tuning the ratio of uneven coupling strength.  First a qubit is prepared at ground state $\psi(0) = |n,a_1\big\rangle$.  Given time, it will evolve accordingly,
\begin{eqnarray}\label{eqn:psi}
 \psi(t) &=& (2\sqrt{1+\delta^2})^{-1}\Big( |E_+\big\rangle e^{-iE_+ t} - |E_-\big\rangle e^{-iE_- t} \Big) \nonumber\\
 &=& e^{-in\omega t} \Big\{ cos\Omega t |n,a_1\big\rangle - i sin\Omega t \frac{\delta}{\sqrt{1+\delta^2}}\Big( \delta |n-1,b_1\big\rangle + |n-1,b_2\big\rangle \Big) \Big\}, 
\end{eqnarray}
where we regard the coupling strength ratio $\delta=g_{11}/g_{12}$ as a controllable parameter.  We remark that at late time $t_r=\pi/2\Omega$, a superposition of excited states $\psi(t_r)$ is created with coefficients tuned by $\delta$.  This suggests that the coupling strength should be dressed with angular dependence, which might be closely related to the BMS transformation of soft hairs.  With that being said, the analytic continuation $\Omega \to i\Omega^\prime$ will bring equation (\ref{eqn:psi}) to following form
\begin{equation}\label{eqn:psi_2}
 \psi(t) \sim \frac{1}{2\sqrt{1+\delta^2}}\Big(  |E_+\big\rangle e^{\Omega^\prime t} - |E_-\big\rangle e^{-\Omega^\prime t} \Big),   
\end{equation}
in comparison to the Goldstone boson mode \footnote{In \cite{Maitra:2019eix} the black hole metric adopts ingoing Eddington-Finkelstein coordinates such that $\nu=t+r*$.  To show its similarity to (\ref{eqn:psi_2}), we have promoted the classical boson field $F$ to operator and substitute $A$ and $B$ with $A^{+}$ and $-A^{-}$ respectively.}
\begin{equation}
    \hat{F}(\nu,\theta,\phi) \sim \sum_{\l m} c_{\l m}Y_{\l m}(\theta,\phi) \Big( \hat{A}_{+} e^{\tilde{\Omega}\nu} - \hat{A}_{-} e^{-\tilde{\Omega}\nu}\Big)
\end{equation}
where both $\Omega^\prime$ and $\tilde{\Omega}$ are of order $1/M$.  Assuming there exists a black hole state $|BH\big\rangle$ such that $\hat{A}_{\pm}|BH\big\rangle \sim |E_{\pm}\big\rangle$.  This similarity indeed implies that $\delta$ is a function of angular dependence.

At last, we may label those degenerate black hole states $|\hat{m}_i,\hat{\theta}_i,\hat{\phi}_i \big\rangle$ by mutually commuting quantum numbers. Let $\hat{m}_i$ denotes the {\sl principle} quantum number, which is given by quantized black hole mass in terms of Planck mass, i.e. $\hat{m}_i=M_i/m_{p}$.  $\hat{\theta}_i$ and $\hat{\phi}_i$ are quantized spherical angles such that $\Delta \hat{\theta}_i = \Delta \hat{\phi}_i = c^2\l_{p}/(2GM_i)$. Then the interaction terms in two-level Hamiltonian (\ref{eqn:hamiltonian}) are refined as
\begin{eqnarray}\label{eqn:hamiltonian_2}
\cdots &+&  \sum_{a_i} \sum_{b_j} g_{{a_i}{b_j}}(\hat{\theta}_{a_i},\hat{\phi}_{a_i},\hat{\theta}_{b_j},\hat{\phi}_{b_j}) \big( \hat{\alpha}|\hat{m}_{b_j},\hat{\theta}_{b_j},\hat{\phi}_{b_j} \big\rangle \big \langle |\hat{m}_{a_i},\hat{\theta}_{a_i},\hat{\phi}_{a_i} | + \hat{\alpha}^{\dagger} |\hat{m}_{a_i},\hat{\theta}_{a_i},\hat{\phi}_{a_i} \big\rangle \big \langle |\hat{m}_{b_j},\hat{\theta}_{b_j},\hat{\phi}_{b_j} | \big)  \nonumber\\
&+& \sum_{a_i} \sum_{a_k} J_{a_{i}a_{k}}(\hat{\theta}_{a_i},\hat{\phi}_{a_i},\hat{\theta}_{a_k},\hat{\phi}_{a_k})
\big( |\hat{m}_{a_i},\hat{\theta}_{a_i},\hat{\phi}_{a_i} \big\rangle \big \langle |\hat{m}_{a_k},\hat{\theta}_{a_k},\hat{\phi}_{a_k} | + |\hat{m}_{a_i},\hat{\theta}_{a_i},\hat{\phi}_{a_i} \big\rangle \big \langle |\hat{m}_{a_k},\hat{\theta}_{a_k},\hat{\phi}_{a_k} | \big)  \nonumber\\
&+& \sum_{b_j} \sum_{b_l} J_{b_{j}b_{l}}(\hat{\theta}_{b_j},\hat{\phi}_{b_j},\hat{\theta}_{b_l},\hat{\phi}_{b_l})
\big( |\hat{m}_{b_j},\hat{\theta}_{b_j},\hat{\phi}_{b_j} \big\rangle \big \langle |\hat{m}_{b_l},\hat{\theta}_{b_l},\hat{\phi}_{b_l} | + |\hat{m}_{b_l},\hat{\theta}_{b_l},\hat{\phi}_{b_l} \big\rangle \big \langle |\hat{m}_{b_j},\hat{\theta}_{b_j},\hat{\phi}_{b_j} | \big) 
\end{eqnarray}
for angular dependent photon coupling $g_{ij}$ (among different mass/energy levels) and hopping coupling $J_{ij}$ (among same mass/energy levels).

\section{Degree of Entanglement}
On the other hand, given a mixed excited state such as that in the previous section, one is interested in the time evolution of its degree of entanglement.  In this letter, we adopt the mutual entropy method (DEM) proposed in \cite{Belavkin:1998vu} and its application to JC model \cite{Furuichi:1999hy}.  We prepare the initial state 
\begin{equation}
    \rho(0) = \lambda_1  |n-1,b_1\big\rangle \big\langle n-1,b_1| + \lambda_2 |n-1,b_2\big\rangle \big\langle n-1,b_2|
\end{equation}
and photons are in a coherent state
\begin{equation}
    \omega = |\theta\big\rangle \big\langle \theta |, \quad |\theta\big\rangle = \exp\Big( -\frac{1}{2}|\theta|^2\Big)\sum_j \frac{\theta^j}{\sqrt{j!}}|j\big\rangle ,
\end{equation}
\begin{figure}[t]
\includegraphics{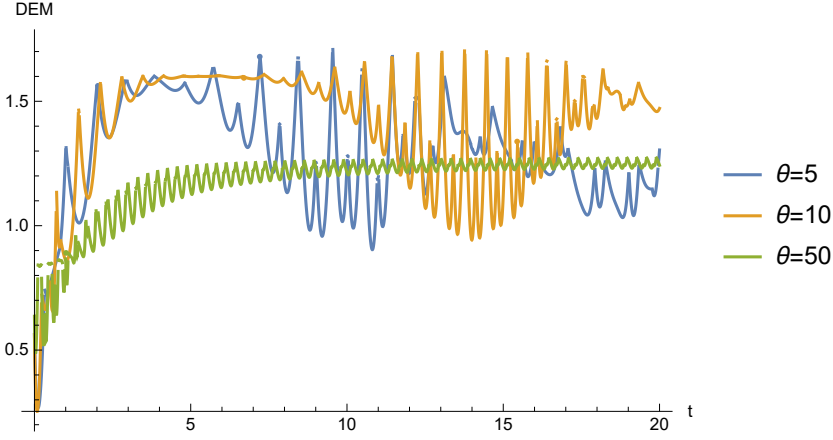}
\caption{\label{fig:DEM} Evolution of degree of entanglement due to mutual entropy (DEM) in our toy model.  We plot for different average photon numbers $\theta=5, 10, 50$ given $\lambda_1=0.25, \lambda_2=0.75$.  The time evolution shows periodic Rabi oscillation as usual JC model.  For more photons in the cavity, the model achieves its strongest entanglement at later time but also lasts longer.}
\end{figure}

The time evolution for the JC model between black atom and coherent field is generated by $\hat{{\cal H}}$, namely, given the unitary operator $U_t = \exp(-it\hat{{\cal H}})$, we have 
\begin{eqnarray}
    \rho(t) &=& U_t \Big( \rho(0)\otimes\omega \Big) U_t^* \nonumber\\
    &=& c_{11}(t) |n,a_1\big\rangle \big\langle n,a_1| + c_{12}(t) |n,a_1\big\rangle \big\langle n-1,b_1| + c_{13}(t) |n,a_1\big\rangle \big\langle n-1,b_2| \nonumber\\
    &+& c_{21}(t) |n-1,b_1\big\rangle \big\langle n,a_1| + c_{22}(t) |n-1,b_1\big\rangle \big\langle n-1,b_1| + c_{23}(t) |n-1,b_1\big\rangle \big\langle n-1,b_2| \nonumber\\
    &+& c_{31}(t) |n-1,b_2\big\rangle \big\langle n,a_1| + c_{32}(t) |n-1,b_2\big\rangle \big\langle n-1,b_1| + c_{33}(t) |n-1,b_2\big\rangle \big\langle n-1,b_2|  \nonumber
\end{eqnarray}
where 
\begin{eqnarray}
c_{11} &=& \frac{1}{2}\exp\big(-|\theta|^2\big)\big(\lambda_1+\lambda_2\big)\sum_n \frac{|\theta|^{2n}}{n!}\sin^2\Omega t \nonumber\\
c_{22} &=& \frac{1}{4}\exp\big(-|\theta|^2\big)\sum_n \frac{|\theta|^{2n}}{n!}\big\{\lambda_1  (1+\cos \Omega t)^2+\lambda_2 (\cos \Omega t-1)^2\big\} = c_{33} \nonumber\\
c_{12} &=& \frac{-i}{2\sqrt{2}}\exp\big(-|\theta|^2\big)\sum_n \frac{|\theta|^{2n}}{n!}\sin\Omega t\big\{\lambda_1  (1+\cos \Omega t)+\lambda_2 (\cos \Omega t-1)\big\} = c_{21}^* \nonumber\\
c_{13} &=& \frac{-i}{2\sqrt{2}}\exp\big(-|\theta|^2\big)\sum_n \frac{|\theta|^{2n}}{n!}\sin\Omega t\big\{\lambda_1  (\cos \Omega t-1)+\lambda_2 (1+\cos \Omega t)\big\} = c_{31}^*\nonumber\\
c_{23} &=& \frac{1}{4}\exp\big(-|\theta|^2\big)\sum_n \frac{|\theta|^{2n}}{n!}(-\sin^2 \Omega t)\big\{\lambda_1 +\lambda_2\big\} = c_{32}^* \nonumber
\end{eqnarray}
Following \cite{Furuichi:1999hy}, one can obtain the reduced density operator $\rho(t)^A$ and $\rho(t)^F$ for black atom and radiation field respectively.  Then the DEM is computed as
\begin{eqnarray}
 {\cal I}_\rho(\rho(t)^A,\rho(t)^F) &=& -c_{11}\log c_{11} - c_{22}\log c_{22} - c_{33}\log c_{33} \\
 &+& c_{12}\log c_{12} + c_{21}\log c_{21} + c_{23}\log c_{23} + c_{32}\log c_{32} + c_{13}\log c_{13} + c_{31}\log c_{31} \nonumber
\end{eqnarray}
In the Fig. \ref{fig:DEM}, we show time evolution of degree of entanglement due to mutual entropy (DEM) in our toy model.  We plot for different average photon numbers $\theta=5, 10, 50$ given $\lambda_1=0.25$ and $\lambda_2=0.75$.  The time evolution shows periodic Rabi oscillation as usual JC model.  For more photons in the cavity, the model achieves its strongest entanglement at slightly later time but the entanglement lasts much longer.  Instead of our thought experiment, which concerns black holes in a very large cavity, one may consider a single black hole surrounded by resonant photons trapped inside the photon sphere, which behaves like a cavity with semi-transparent walls.  Our simulation suggests that a black hole can retain its information by sustaining entanglement for a long time since it is probably surrounded by huge amount of photons.

\section{Conclusion and discussions}
In this letter, we construct a toy JC model of black hole and regard the Hawking radiation as the stimulated emission.  We conclude that  doublet microstates are seated at event horizon.  While those degrees of freedom might be the soft gravitons in the BMS transformation, there could be other candidates such as the spin foam in the Loop Quantum Gravity \cite{Reisenberger:1996pu, Baez:1997zt}.  We further argue that coupling strength could  have angular dependence in parallel to those arbitrary functions appeared in the components of Goldstone boson modes.  Later we also show this angular feature can be woven into entanglement among excited states.  In general, those eigenstates in (\ref{eqn:psi_2}) should be given more angular quantum numbers like $|E^{lm}_\pm \big\rangle$ if the complete interaction (\ref{eqn:hamiltonian_2}) is considered.  Then the total $N\sim e^{M^2/l^2_p}$ coefficients in front of $|E^{lm}_\pm \big\rangle$ can be uniquely determined by the same number of coefficient $c_{lm}$ in front of Goldstone boson modes.  At last, our JC model illustrates that a entangled system may still stay coherent for a long while provided interaction with large amount of photons.  It also needs to point out that the degree of entanglement oscillates at Rabi frequency as expected in any JC model.  We argue that all these features could be seen in a realistic black hole system where the photon sphere might have played the role of cavity wall.  As a final remark, we have seen our quantum mechanical model of black holes can capture several features of nonthermal Hawking radiation and black hole information. It is interesting to see how far one can model black holes without spacetime and to which stage spacetime curvature does concern.  We will leave them for future study.
 
\begin{acknowledgments}
We would like to thank Long-Ke Lin at CYCU for discussion on JC model.  This work is supported in part by the Taiwan's Ministry of Science and Technology (grant No. 106-2112-M-033-007-MY3) and the National Center for Theoretical Sciences (NCTS).
\end{acknowledgments}

\appendix

\section{The Jaynes-Cummings model}
The Jaynes-Cummings model for the black hole atom in our thought experiment is proposed as follows:
 \begin{equation}\label{eqn:hamiltonian_3}
  \hat{{\cal H}} = \sum_i M_a |a_i\big \rangle \big \langle a_i| + \sum_j M_b |b_j\big \rangle \big \langle b_j| 
  +  \omega \hat{\alpha}^{\dagger}\hat{\alpha} 
  +  \sum_i \sum_j  g_{ij} \big( \hat{\alpha}|b_j\big \rangle \big \langle a_i| + \hat{\alpha}^{\dagger} |a_i\big \rangle \big \langle b_j| \big) 
 \end{equation}

where indices $i,j$ label each degeneracy state.  $\hat{\alpha}$ and $\hat{\alpha}^{\dagger}$ are annihilation and creation operators of photons.  The couplings $g_{ij}$ are responsible for emission and absorption.  Without loss of generality, we only focus on transition among just few states, say $|n,a_1\big \rangle$ for the ground state, and $|n-1,b_1\big \rangle$, $|n-1,b_2\big \rangle$ for two degenerate excited states.  We define the product state, for instance, $|n,a_1\big \rangle = |n\big \rangle \otimes |a_1\big \rangle$, where $n$ is the photon occupation number in the cavity and spin index $s$ is suppressed in atom states.  The Hamiltonian in its matrix form reads
\[
\begin{bmatrix}
\big\langle n,a_1 |\hat{{\cal H}}| n,a_1 \big\rangle & \big\langle n,a_1 |\hat{{\cal H}}| n-1,b_1 \big\rangle &
\big\langle n,b_1 |\hat{{\cal H}}| n-1,b_2 \big\rangle \\
\big\langle n-1,b_1 |\hat{{\cal H}}| n,a_1 \big\rangle & \big\langle n-1,b_1 |\hat{{\cal H}}| n-1,b_1 \big\rangle &
\big\langle n-1,b_1 |\hat{{\cal H}}| n-1,b_2 \big\rangle \\
\big\langle n-1,b_2 |\hat{{\cal H}}| n,a_1 \big\rangle & \big\langle n-1,b_2 |\hat{{\cal H}}| n-1,b_1 \big\rangle &
\big\langle n-1,b_2 |\hat{{\cal H}}| n-1,b_2 \big\rangle \\
\end{bmatrix}
\]
The eigenvalues then read
\begin{equation}
E_0 = n\omega + \Delta, \quad E_{\pm} = n\omega + \frac{\Delta}{2} \pm \sqrt{(g_{11}^2+g_{12}^2) n +\Delta^2/4},
\end{equation}
where the detuning parameter $\Delta \equiv M_b - M_a -\omega $.  The corresponding eigenstates (before normalization) are 
\begin{eqnarray}
&& |E_0\big \rangle = -\frac{g_{12}}{g_{11}}|n-1,b_1\big\rangle + |n-1,b_2\big \rangle ,\\
&& |E_\pm\big \rangle = \Big(-\frac{\Delta}{2g_{12}\sqrt{n}}\pm \sqrt{\big(\frac{g_{11}}{g_{12}}\big)^2 + 1 + \big(\frac{\Delta}{2g_{12}\sqrt{n}}\big)^2}  \Big) |n,a_1\big\rangle +\frac{g_{11}}{g_{12}} |n-1,b_1\big\rangle + |n-1,b_2\big\rangle.\nonumber
\end{eqnarray}
In particular, at the resonance $\Delta=0$ and even coupling strength $g_{11}=g_{12}=g$, one obtains the time evolution of states:
\begin{eqnarray}\label{eqn:time_evolution}
 |n,a_1\big\rangle_t &=&  \cos \Omega t|n, a_1\big\rangle_0  +\frac{i}{\sqrt{2}}\sin \Omega t\Big( |n-1,b_1\big\rangle_0 + |n-1,b_2\big\rangle_0 \Big)  , \nonumber\\
 |n-1,b_1\big\rangle_t &=& \frac{i}{\sqrt{2}}\sin\Omega t|n, a_1\big\rangle_0  + \Big(\frac{1}{2\sqrt{2}}+\frac{1}{2}\cos\Omega t \Big)|n-1,b_1\big\rangle_0 + \nonumber\\
 &&\Big(-\frac{1}{2\sqrt{2}}+\frac{1}{2}\cos\Omega t \Big)|n-1,b_2\big\rangle_0 , \nonumber\\
 |n-1,b_2\big\rangle_t &=& \frac{i}{\sqrt{2}}\sin\Omega t|n, a_1\big\rangle_0  + \Big(-\frac{1}{2\sqrt{2}}+\frac{1}{2}\cos\Omega t \Big)|n-1,b_1\big\rangle_0 + \nonumber\\
 &&\Big(\frac{1}{2\sqrt{2}}+\frac{1}{2}\cos\Omega t \Big)|n-1,b_2\big\rangle_0 ,
\end{eqnarray}
for given initial states $|n,a_1\big\rangle_0$, $|n-1,b_1\big\rangle_0$ and $|n-1,b_2\big\rangle_0$.  Here we ignore the overall phase factor $e^{in\omega t}$ and denote $\Omega \equiv \sqrt{2g^2n}$.
One can further obtain the unitary operator for time evolution
\begin{eqnarray}
\exp(-it\hat{{\cal H}}) &=& |0,a_1\big\rangle \big\langle 0,a_1| + \sum_n^{\infty} e^{-in\omega t}\Big\{ \cos \Omega t |n,a_1\big\rangle \big\langle n,a_1| \nonumber\\
&+& \frac{1}{\sqrt{2}}\big(-i\sin \Omega t \big) \big(|n,a_1\big\rangle \big\langle n-1,b_1|
+ |n-1,b_1\big\rangle \big\langle n,a_1|\big) \nonumber\\ 
&+&   \frac{1}{\sqrt{2}}\big(-i\sin \Omega t \big) \big(|n,a_1\big\rangle \big\langle n-1,b_2| + |n-1,b_2\big\rangle \big\langle n,a_1|\big) \nonumber\\
&+& \frac{1}{2}\big( 1+\cos \Omega t \big) |n-1,b_1\big\rangle \big\langle n-1,b_1| \nonumber\\
&+& \frac{1}{2}\big(\cos \Omega t-1 \big) \big(|n-1,b_1\big\rangle \big\langle n-1,b_2| + |n-1,b_2\big\rangle \big\langle n-1,b_1| \big) \nonumber\\
&+& \frac{1}{2}\big( 1+\cos \Omega t \big) |n-1,b_2\big\rangle \big\langle n-1,b_2| \Big\}
\end{eqnarray}


\begin{thebibliography}{99}

%\cite{Parikh:1999mf}
\bibitem{Parikh:1999mf} 
  M.~K.~Parikh and F.~Wilczek,
  ``Hawking radiation as tunneling,''
  Phys.\ Rev.\ Lett.\  {\bf 85}, 5042 (2000)
  doi:10.1103/PhysRevLett.85.5042
  [hep-th/9907001].
  %%CITATION = doi:10.1103/PhysRevLett.85.5042;%%
  %1285 citations counted in INSPIRE as of 11 Oct 2019

%\cite{Zhang:2009jn}
\bibitem{Zhang:2009jn} 
  B.~Zhang, Q.~y.~Cai, L.~You and M.~s.~Zhan,
  ``Hidden Messenger Revealed in Hawking Radiation: A Resolution to the Paradox of Black Hole Information Loss,''
  Phys.\ Lett.\ B {\bf 675}, 98 (2009)
  doi:10.1016/j.physletb.2009.03.082
  [arXiv:0903.0893 [hep-th]].
  %%CITATION = doi:10.1016/j.physletb.2009.03.082;%%
  %88 citations counted in INSPIRE as of 11 Oct 2019

\bibitem{Einstein:1916}
  A.~Einstein, "Strahlungs-emission und -absorption nach der Quantentheorie," Verhandlungen der Deutschen Physikalischen Gesellschaft {\bf 18}, 318-323 (1916).

\bibitem{Bondi:1962}
Bondi, H.. 1962. M. G. J. van der Burg and A. W. K. Metzner, Gravitational waves in general relativity. 7. Waves from axisymmetric isolated systems. Proc.Roy.Soc.Lond.,269,21 

%\cite{Sachs:1962zza}
\bibitem{Sachs:1962zza} 
  R.~Sachs,
  ``Asymptotic symmetries in gravitational theory,''
  Phys.\ Rev.\  {\bf 128}, 2851 (1962).
  doi:10.1103/PhysRev.128.2851
  %%CITATION = doi:10.1103/PhysRev.128.2851;%%
  %382 citations counted in INSPIRE as of 09 Aug 2019
  
\bibitem{Sachs:1962}
Sachs, R.K.. 1962. Gravitational waves in general relativity. 8. Waves in asymptotically flat space-times. Proc.Roy.Soc.Lond.,270,103

%\cite{Dvali:2011aa}
\bibitem{Dvali:2011aa} 
  G.~Dvali and C.~Gomez,
  ``Black Hole's Quantum N-Portrait,''
  Fortsch.\ Phys.\  {\bf 61}, 742 (2013)
  doi:10.1002/prop.201300001
  [arXiv:1112.3359 [hep-th]].
  %%CITATION = doi:10.1002/prop.201300001;%%
  %235 citations counted in INSPIRE as of 09 Aug 2019

%\cite{Dvali:2015ywa}
\bibitem{Dvali:2015ywa} 
  G.~Dvali, A.~Franca, C.~Gomez and N.~Wintergerst,
  ``Nambu-Goldstone Effective Theory of Information at Quantum Criticality,''
  Phys.\ Rev.\ D {\bf 92}, no. 12, 125002 (2015)
  doi:10.1103/PhysRevD.92.125002
  [arXiv:1507.02948 [hep-th]].
  %%CITATION = doi:10.1103/PhysRevD.92.125002;%%
  %24 citations counted in INSPIRE as of 09 Aug 2019
  
%\cite{Dvali:2015rea}
\bibitem{Dvali:2015rea} 
  G.~Dvali, C.~Gomez and D.~Lüst,
  ``Classical Limit of Black Hole Quantum N-Portrait and BMS Symmetry,''
  Phys.\ Lett.\ B {\bf 753}, 173 (2016)
  doi:10.1016/j.physletb.2015.11.073
  [arXiv:1509.02114 [hep-th]].
  %%CITATION = doi:10.1016/j.physletb.2015.11.073;%%
  %37 citations counted in INSPIRE as of 09 Aug 2019
 
%\cite{Hawking:2016msc}
\bibitem{Hawking:2016msc} 
  S.~W.~Hawking, M.~J.~Perry and A.~Strominger,
  ``Soft Hair on Black Holes,''
  Phys.\ Rev.\ Lett.\  {\bf 116}, no. 23, 231301 (2016)
  doi:10.1103/PhysRevLett.116.231301
  [arXiv:1601.00921 [hep-th]].
  %%CITATION = doi:10.1103/PhysRevLett.116.231301;%%
  %344 citations counted in INSPIRE as of 09 Aug 2019

%\cite{Strominger:2017aeh}
\bibitem{Strominger:2017aeh} 
  A.~Strominger,
  ``Black Hole Information Revisited,''
  arXiv:1706.07143 [hep-th].
  %%CITATION = ARXIV:1706.07143;%%
  %42 citations counted in INSPIRE as of 12 Aug 2019
  
%\cite{Chu:2018tzu}
\bibitem{Chu:2018tzu} 
  C.~S.~Chu and Y.~Koyama,
  ``Soft Hair of Dynamical Black Hole and Hawking Radiation,''
  JHEP {\bf 1804}, 056 (2018)
  doi:10.1007/JHEP04(2018)056
  [arXiv:1801.03658 [hep-th]].
  %%CITATION = doi:10.1007/JHEP04(2018)056;%%
  %6 citations counted in INSPIRE as of 12 Aug 2019

%\cite{Averin:2016hhm}
\bibitem{Averin:2016hhm} 
  A.~Averin, G.~Dvali, C.~Gomez and D.~Lust
  ``Goldstone origin of black hole hair from supertranslations and criticality,''
  Mod.\ Phys.\ Lett.\ A {\bf 31}, no. 39, 1630045 (2016)
  doi:10.1142/S0217732316300457
  [arXiv:1606.06260 [hep-th]].
  %%CITATION = doi:10.1142/S0217732316300457;%%
  %27 citations counted in INSPIRE as of 09 Aug 2019

%\cite{Eling:2016xlx}
\bibitem{Eling:2016xlx} 
  C.~Eling and Y.~Oz,
  ``On the Membrane Paradigm and Spontaneous Breaking of Horizon BMS Symmetries,''
  JHEP {\bf 1607}, 065 (2016)
  doi:10.1007/JHEP07(2016)065
  [arXiv:1605.00183 [hep-th]].
  %%CITATION = doi:10.1007/JHEP07(2016)065;%%
  %34 citations counted in INSPIRE as of 09 Aug 2019

%\cite{Sorkin:2014kta}
\bibitem{Sorkin:2014kta} 
  R.~D.~Sorkin,
  ``1983 paper on entanglement entropy: "On the Entropy of the Vacuum outside a Horizon",''
  arXiv:1402.3589 [gr-qc].
  %%CITATION = ARXIV:1402.3589;%%
  %57 citations counted in INSPIRE as of 09 Aug 2019
  
  %\cite{tHooft:1993dmi}
\bibitem{tHooft:1993dmi} 
  G.~'t Hooft,
  ``Dimensional reduction in quantum gravity,''
  Conf.\ Proc.\ C {\bf 930308}, 284 (1993)
  [gr-qc/9310026].
  %%CITATION = GR-QC/9310026;%%
  %2212 citations counted in INSPIRE as of 10 Aug 2019
  
%\cite{Averin:2016ybl}
\bibitem{Averin:2016ybl} 
  A.~Averin, G.~Dvali, C.~Gomez and D.~Lust,
  ``Gravitational Black Hole Hair from Event Horizon Supertranslations,''
  JHEP {\bf 1606}, 088 (2016)
  doi:10.1007/JHEP06(2016)088
  [arXiv:1601.03725 [hep-th]].
  %%CITATION = doi:10.1007/JHEP06(2016)088;%%
  %49 citations counted in INSPIRE as of 09 Aug 2019

\bibitem{Jaynes:1963}
  E.~T.~Jaynes and F.~W.~Cummings,
  ''Comparison of quantum and semiclassical radiation theories with application to the beam maser.'' Proceedings of the IEEE, {\bf 51}, 89 (1963)

%\cite{Braunstein:2011gz}
\bibitem{Braunstein:2011gz} 
  S.~L.~Braunstein and M.~K.~Patra,
  ``Black hole evaporation rates without spacetime,''
  Phys.\ Rev.\ Lett.\  {\bf 107}, 071302 (2011)
  doi:10.1103/PhysRevLett.107.071302
  [arXiv:1102.2326 [quant-ph]].
  %%CITATION = doi:10.1103/PhysRevLett.107.071302;%%
  %14 citations counted in INSPIRE as of 11 Oct 2019

%\cite{Clerk:2010}
\bibitem{Clerk:2010}
A.~A.~Clerk, M.~H.~Devoret, S.~M.~Girvin, Florian Marquardt and R.~J.~Schoelkopf, ``Introduction to quantum noise, measurement, and amplification,''
Rev.\ Mod.\ Phys.\ {\bf 82}, 1155 (2010)

%\cite{Maitra:2019eix}
\bibitem{Maitra:2019eix} 
  M.~Maitra, D.~Maity and B.~R.~Majhi,
  ``Near horizon symmetries, emergence of Goldstone modes and thermality,''
  arXiv:1906.04489 [hep-th].
  %%CITATION = ARXIV:1906.04489;%%
  
%\cite{Wen:2014xpa}
\bibitem{Wen:2014xpa} 
  W.~Y.~Wen,
  ``Nonthermal correction to black hole spectroscopy,''
  Eur.\ Phys.\ J.\ C {\bf 75}, no. 2, 78 (2015)
  doi:10.1140/epjc/s10052-015-3302-3
  [arXiv:1411.3999 [gr-qc]].
  %%CITATION = doi:10.1140/epjc/s10052-015-3302-3;%%
  %1 citations counted in INSPIRE as of 12 Aug 2019


%\cite{Belavkin:1998vu}
\bibitem{Belavkin:1998vu} 
  V.~P.~Belavkin and M.~Ohya,
  ``Quantum entanglements and entangled mutual entropy,''
  Proc.\ Roy.\ Soc.\ Lond.\ A {\bf 458}, 209 (2001)
  [quant-ph/9812082].
  %%CITATION = QUANT-PH/9812082;%%
  %3 citations counted in INSPIRE as of 05 Oct 2019

%\cite{Furuichi:1999hy}
\bibitem{Furuichi:1999hy}
  S.~Furuichi and M.~Ohya,
  ``Entanglement degree for jaynes-cummings model,''
  Lett.\ Math.\ Phys.\  {\bf 49} (1999) 279
  doi:10.1023/A:1007684527101
  [quant-ph/9903004].
  %%CITATION = doi:10.1023/A:1007684527101;%%
  %2 citations counted in INSPIRE as of 05 Oct 2019
  
%\cite{Reisenberger:1996pu}
\bibitem{Reisenberger:1996pu} 
  M.~P.~Reisenberger and C.~Rovelli,
  %``'Sum over surfaces' form of loop quantum gravity,''
  Phys.\ Rev.\ D {\bf 56}, 3490 (1997)
  doi:10.1103/PhysRevD.56.3490
  [gr-qc/9612035].
  %%CITATION = doi:10.1103/PhysRevD.56.3490;%%
  %278 citations counted in INSPIRE as of 14 Oct 2019
  
%\cite{Baez:1997zt}
\bibitem{Baez:1997zt} 
  J.~C.~Baez,
  %``Spin foam models,''
  Class.\ Quant.\ Grav.\  {\bf 15}, 1827 (1998)
  doi:10.1088/0264-9381/15/7/004
  [gr-qc/9709052].
  %%CITATION = doi:10.1088/0264-9381/15/7/004;%%
  %326 citations counted in INSPIRE as of 14 Oct 2019

\end{thebibliography}
\end{document}